\documentclass[12pt]{preprint}
\usepackage[dvips]{graphicx}
\usepackage{amsmath}
\usepackage{amsfonts}
\usepackage{amssymb}
\usepackage{times}
\usepackage{sub_JP}
\usepackage{MHequ}

\def\OFR{{$\Omega$-FR }}

\def\LFR{{$\Lambda$-FR }}
\def\WFR{{$\Omega$-FR }}

\def\H{{\rm H}}
\def\C{{\rm C}}
\def\L{{\rm C}}
\def\R{{\rm H}}

\def\kB{k_{\rm B}}
\def\eref#1{(\ref{#1})}
\let\rho=\varrho

\def\NH{{Nos\'e-Hoover }}
\def\dof{{\em d.o.f. }}
\def\nth{n_{\rm th}}
\def\tnh{\tau_{\rm th}}
\def\xiL{\xi_{\rm C}}
\def\xiR{\xi_{\rm H}}

\newcommand{\zG}{\Gamma}
\newcommand{\zW}{\Omega}
\newcommand{\zL}{\Lambda}

\newtheorem{theorem}{Theorem}[section]

\newtheorem{remark}[theorem]{Remark}

\begin{document}
\title{On  the  Fluctuation Relation for Nos\'e-Hoover Boundary  Thermostated
  Systems}
\author{Carlos Mej\'{\i}a-Monasterio, Lamberto Rondoni}
\institute{Dipartimento de Matematica, Politecnico di Torino, corso Duca degli  Abruzzi 24, 10129 Torino Italy}
\maketitle

\begin{abstract}
  We  discuss  the  transient  and  steady state fluctuation  relation  for  a
  mechanical system in contact with two deterministic thermostats at different
  temperatures.   The system  is a  modified Lorentz  gas in  which  the fixed
  scatterers exchange  energy with the  gas of particles, and  the thermostats
  are modelled by  two Nos\'e-Hoover thermostats applied at  the boundaries of
  the  system.  The  transient fluctuation  relation, which  holds only  for a
  precise  choice  of the  initial  ensemble, is  verified  at  all times,  as
  expected.   Times  longer  than  the  mesoscopic  scale,  needed  for  local
  equilibrium to be  settled, are required if a  different initial ensemble is
  considered.    This   shows   how   the   transient   fluctuation   relation
  asymptotically  leads  to the  steady  state  relation  when, as  explicitly
  checked in our systems, the condition found in [D.J.\ Searles, {\em et al.},
  J.\ Stat.\  Phys.  128, 1337 (2007)],  for the validity of  the steady state
  fluctuation relation, is verified.  For the steady state fluctuations of the
  phase space contraction rate $\zL$  and of the dissipation function $\zW$, a
  similar relaxation regime at shorter averaging times is found.  The quantity
  $\zW$ satisfies with good accuracy the fluctuation relation for times larger
  than  the mesoscopic  time  scale; the  quantity  $\zL$ appears  to begin  a
  monotonic convergence  after such times.   This is consistent with  the fact
  that $\zW$ and  $\zL$ differ by a total time derivative,  and that the tails
  of the probability distribution function of $\zL$ are Gaussian.
\end{abstract}
\thispagestyle{empty}

\section{Introduction}
\label{sec:intro}

In  the  last  decades it  has  been  recognized  the  central role  that  the
fluctuations  around  stationary  states  play  for a  microscopic  theory  of
nonequilibrium  phenomena \cite{RMM}.   Furthermore,  biological sciences  and
nano-technology are ever more interested in nonequilibrium fluctuations.

The 1993 paper by Evans, Cohen  and Morriss \cite{ECM}, on the fluctuations of
the entropy  production rate  of a deterministic  particle system,  modeling a
shearing fluid, provided a unifying  framework for a variety of nonequilibrium
phenomena, under  a mathematical  expression nowadays called  {\em Fluctuation
  Relation} (FR). They proposed and tested the following formula
\begin{equ} \label{firstFR}
\frac{P(\overline{A}_\tau \in A_\delta)}{P(\overline{A}_\tau \in -A_\delta)}
\simeq e^{\tau  A} 
\end{equ}
where $\overline{A}_\tau$ is the average  of the power dissipated on evolution
segments of  duration $\tau$, and  $P(\overline{A}_\tau \in \pm  A_\delta)$ is
the  probability that  $\overline{A}_\tau$ takes  a  value in  an interval  of
length  $\delta$, centered at  $\pm A$.   In analogy  with the  periodic orbit
expansions \cite{POEinECM2a,POEinECM2b},  Eq.\eref{firstFR}, was obtained from
the   ``Lyapunov   weights''   in   the  long   $\tau$   limit.    Remarkably,
Eq.~\eref{firstFR}, does not contain any adjustable parameter.

In 1994,  Evans and Searles  obtained the first  transient FRs for  the energy
dissipation    rate    divided   by    $k_B    T$,    denoted   by    $\Omega$
\cite{review,earlierpapersA,earlierpapersB,generalized,SE2000,StephenPRE,stephennew},
which we call transient \OFR, and  which concern the statistics of an evolving
ensemble  of  systems, instead  of  the  steady  state statistics.   The  only
requirement  for the  transient \OFR's  to hold  is the  reversibility  of the
microscopic dynamics.  Because they  describe the behaviour of $\Omega$, these
relations can be experimentally verified \cite{WSMSE}.

In 1995,  Gallavotti and  Cohen provided a  mathematical justification  of the
Lyapunov  weights  of   Ref.\cite{ECM},  introducing  the  Chaotic  Hypothesis
\cite{GCa,GCb,GG-MPEJ}, which states that: ``A reversible many-particle system
in  a  stationary  state  can  be  regarded  as  a  transitive  Anosov  system,
\footnote{Anosov  systems  are  smooth  and uniformly  hyperbolic.}   for  the
purpose of  computing its macroscopic  properties''. The result was  a genuine
steady state  FR, which we call \LFR,  as it concerns the  fluctuations of the
phase  space contraction  rate  $\Lambda$.  This  quantity  equals the  energy
dissipation  rate  $\zW$  in  a  subclass of  Gaussian  isoenergetic  particle
systems, which  includes the  model of \cite{ECM},  while in many  other cases
$\zL$ differs from $\zW$ by a total  derivative.  As far as we know, the works
\cite{GCa,GCb,GG-MPEJ} provide the  only answer that has been  given so far to
the question  of which dynamical  systems can be  proven to verify  the steady
state  $\zL$-FR.  A  strong assumption  as the  Chaotic Hypothesis  raises the
question  of which  systems of  physical interest  are  ``Anosov-like'', since
almost none of them is  actually Anosov.  The answer of Refs.\cite{GCa,GCb} is
that the  Anosov property,  in analogy with  the Ergodic property,  holds ``in
practice'' for sufficiently chaotic, reversible systems.

A  complementary approach,  which  addresses a  different  question, has  been
recently proposed  in \cite{SRE}, developing  ideas first introduced  by Evans
and  Searles, see  e.g.\  \cite{review}. The  question  concerns the  physical
mechanisms at  work in systems which  do obey the steady  state $\zW$-FR. This
approach  leads to  a general  framework  within which  various transient  and
steady  state relations  can be  obtained for  a variety  of  observables, and
especially leads to the identification  of physical mechanisms and time scales
underlying the validity of the steady state FRs. The result is that, while the
transient  relations  only  rest  on  time  reversibility,  the  steady  state
relations do  require further  properties to hold.\footnote{For  instance, the
systems  under  consideration  must  converge  towards a  steady  state.}   In
\cite{SRE}, a property called ``$\Omega$-autocorrelation decay'' is identified
as the  reason for  the steady state  \WFR, as  well as of  a number  of other
asymptotic relations,  to be verified, when  they do.  This  approach leads to
the identification of the time scales  concerning the FRs with the decay times
of a  quantity which we  introduce in Section \ref{sec:SSFT}.   Therefore, the
need arises  to test on concrete  systems the validity  of this identification
and to quantify the corresponding time scales.\footnote{A similar study of the
time scales  was performed in  Ref.\cite{romans}, in one investigation  on the
applicability of  the Chaotic Hypothesis.}   These issues are  closely related
with  the role  played  in the  dynamics  by the  singularities  of the  total
derivative which distinguishes  $\zL$ from $\zW$.  This has  been discussed in
various  papers,  like  \cite{VZC,ESR,BGGZ,Zamp-rev}.   The present  paper  is
devoted to the investigation of such questions.

We study the  heat transport and its transient  and steady state fluctuations,
in a mechanical dissipative system that is maintained out of equilibrium by an
imposed  temperature  gradient.   This  local thermostating  mechanism  leaves
unaltered the Newtonian  dynamics in the bulk of  the system, dissipating only
at its boundaries.   However, due to the interactions,  dissipation may depend
on the non thermostated degrees of freedom ({\it d.o.f.}) as well \cite{loss}.
To impose  a temperature gradient  we use deterministic \NH  thermostats, that
add a time-dependent friction term to  the Hamiltonian of the system in such a
way  that,  if  the  system   is  ergodic,  the  energy  distribution  of  the
thermostated \dof is Boltzmann  distributed with well defined temperature.  As
a consequence of the use of  \NH thermostats, the phase space contraction rate
is  unbounded,  possibly  leading  to  statistical  properties  of  its  large
fluctuations which are not described  by the standard FR (\ref{firstFR}). This
scenario has been considered in various  works, and depends on the form of the
tails  of  the  distribution  of   the  observables  of  interest,  see  e.g.\
\cite{ESR,Zamp-rev,VZC,BGGZ}. As  far as  particle systems are  concerned, the
corresponding modified FR  appears to be consistent with  the \NH thermostated
Lorentz  gas  \cite{gilbert}.   In  our  case,  consistently  with  the  cited
literature, the unboundedness of $\zL$  does not imply any modification of the
standard FR. The distribution of  the fluctuations of $\zL$, indeed, appear to
have Gaussian tails. Therefore,  the search for deterministic particle systems
which verify a modified version of  FR, apart from the case of \cite{gilbert},
must continue.
 
The paper  is organized as  follows: after defining the  many-body dissipative
model that we will use,  in Section~\ref{sec:model} we discuss the equilibrium
and  nonequilibrium  states  that  arise  from the  coupling  with  the  local
thermostats.   In Section~\ref{sec:ft}  we study  the transient  \OFR  and the
steady state fluctuation  relations for the phase space  contraction rate, and
for   the   dissipation   function.    Our  conclusions   are   presented   in
Section~\ref{sec:concl}.

\section{The Model}
\label{sec:model}

The  model  we  consider  consists  of a  gas  of  non-interacting  point-like
particles of mass $m$ that move freely inside a channel
made  of  $L$  identical  two-dimensional  cells.  Each  cell  consists  of  a
rectangular region of height $\Delta y$ and width $\Delta x$, containing two
fixed freely rotating  disks of  radius $R$  and moment  of  inertia $\Theta$.

\begin{figure}[!t]
\begin{center}
  \includegraphics[scale=1.05]{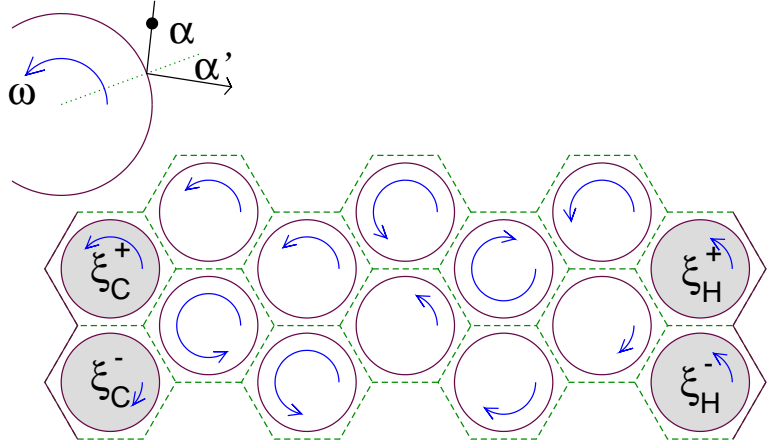}
\caption{Geometry of a system of  length $L=7$ cells. The disks of radius
  $R$ form a  triangular lattice at critical horizon,  meaning that the height
  and width of the cells equal, respectively,  $\Delta x = 2R$ and $\Delta y =
  2W$,  where $W=4R/\sqrt{3}$  is the  separation between  the centers  of the
  disks.   Periodic  boundary  conditions  along the  vertical  direction  are
  considered.  The dashed lines indicate the single hexagonal cells. The solid
  lines  are hard  walls.  At  the  left top  we schematically  show a  single
  collision: when  a particle collides with  a disk at an  angle $\alpha$ with
  respect to  the normal  at the  collision point, the  outgoing angle  of its
  trajectory  $\alpha'$ is determined  by Eq.~\ref{eq:col-rules}.   In between
  collisions, the angular velocity of the disk $\omega$ evolves freely for the
  disks  in  the  bulk  (empty disks)   and  according  to  Eq.~\ref{eq:NH}  for  the
  thermostated disks (filled disks).
\label{fig:model}}
\end{center}
\end{figure}

The position of the disks in each cell and their radius are chosen so that the
geometry of the channel corresponds to  a periodic Lorentz gas in a triangular
lattice  at critical horizon,  defined as  the smallest  disk radius  to disks
separation ratio such that the length of the particle's trajectory between two
collisions with the  disks is bounded.  The lattice of  scatterers is shown in
Fig.~\ref{fig:model}.   Moreover,  the  particles  and disks  are  allowed  to
exchange energy  at collisions,  according to ``perfectly  rough'' collisions,
that  are  reversible, conserve  total  energy  and  angular momentum.   These
collisions are defined by the  following formulae, which relate the normal and
tangential components  of the particle's velocity  ${\mathbf{v}}$ with respect
to  the  disk's surface,  and  the  disk's  angular velocity  $\omega$  before
(unprimed quantities) and after (primed quantities) the collision \cite{mejia}
\begin{equa}[2]\label{eq:col-rules}
&v_n' \ & = & \ \ \ -v_n\, \nonumber\\
&v_t' \ & = & \ \ \  v_t - \frac{2\eta}{1 + \eta} (v_t - R\omega)\,  \\
&R\omega' \ & = & \ \ \ R\omega + \frac{2}{1 + \eta} (v_t - R\omega) \nonumber \ .
\end{equa}
Thus,  collisions are  deterministic, time-reversible  and phase  space volume
preserving. The parameter  $\eta$, defined as 
\begin{equation}
\eta=\frac{\Theta}{mR^2} \ ,
\label{eq:defeta}
\end{equation}
is the only relevant dimensionless parameter characterizing the collision. It
determines the  energy transfer  between disks and  particles in  a collision.
For finite  values of  $\eta$, the particles  indirectly exchange  energy with
each other, through  the interactions with the disks, even  though they do not
directly interact.

This model, introduced in \cite{mejia}  as a mechanical model for coupled heat
and  matter transport,  shows  realistic equilibrium  and  out of  equilibrium
thermodynamical  properties, in  very different  situations, {\em  e.g.}  when
subjected to external gradients of  temperature and chemical potential, in the
presence of external electric or magnetic fields \cite{larralde}.

The simple  energy exchange mechanism  allows the equipartition of  the energy
among  all   the  {\it  d.o.f.},  allowing   the  system  to   reach  a  local
thermodynamical  equilibrium.  In  particular, in  microcanonical simulations,
the particle velocities are  Maxwellian distributed at uniform ``temperature''
consistent   with  the   equipartition  theorem.    These   temperatures  also
characterize  the   distribution  of   angular  velocities  of   the  rotating
scatterers.   The same  is true  in canonical  and  grancanonical simulations,
where the system is subjected to temperature and chemical potential gradients.
One consequence of the establishment  of local thermal equilibrium is that the
gas  of particles  is described  locally by  the equations  of state  of ideal
gases.   Recently, the  properties  of  the stationary  state  for a  modified
version of this model, coupled  to stochastic energy and particles reservoirs,
have  been analytically  obtained in  the weak  coupling limit  \cite{EY}, and
finite coupling corrections have been obtained in \cite{EMMZ}.

\subsection{Local \NH thermostats}
\label{sec:baths}
 
To establish a nonequilibrium state, two heat baths are placed at the left and
right  boundaries of  the model,  with local  temperatures set  to  $T_\L$ and
$T_\R$  respectively.  In  previous  works \cite{mejia,larralde,EY,EMMZ},  the
bulk  of the  system was  coupled to  stochastic thermochemical  baths,  so to
exchange energy and  particles at the boundaries.  Differently,  we consider a
closed  system (particle number  is conserved),  and model  the heat  baths as
local \NH  thermostats \cite{NH}.   Roughly speaking a  \NH thermostat  adds a
kind  of  friction term  to  a  number of  degrees  of  freedom, whose  energy
fluctuations turn to be canonical at some temperature $T$, as a result.

More precisely, consider $\nth$  \dof with generalized positions  $q$, momenta
$\pi$  and  mass  $m$,  subjected  to an  external  potential  $\phi(\vec{q})$
described by the Nos\'e Hamiltonian \cite{nose},
\begin{equa}[3] \label{eq:nose}
&H_{\nth}\left(\vec{q},\vec{\pi},X,\Xi,s\right) & \ = \ &
\sum_{i=1}^{\nth}\frac{\pi_i^2}{2mX^2} + \phi(\vec{q}) +\\
& & \ + \ & \frac{\Xi^2}{2Q} + \left(g+1\right)\kB T\mathrm{ln} X \ .\\
\end{equa}
Here $X$ and $\Xi$ are the generalized position and momentum of the thermostat
and  $Q$ its  effective mass.   $s$ is  a generalized  time variable,  $T$ the
temperature, $g$ a number related to  $\nth$ and $\kB$ the Boltzmann constant. 
If one  identifies $s$ and $\pi$ with  the physical time $t$  and momenta $p$,
letting  $\mathrm{d}s =  X\mathrm{d}t$ and  $\pi_i =  Xp_i$,  and additionally
$g=\nth$  then,  it  is  easy  to  show  that  if  the  generalized  variables
$\vec{\pi}$, $X$ and $\Xi$ are distributed microcanonically, then the physical
variables $\vec{q}$  and $\vec{p}$ are distributed according  to the canonical
distribution with temperature $T$ \cite{evans-book}.

Furthermore, if one is solely  interested in real time averages, the equations
of motion  that derive from  \eref{eq:nose} can be  rewritten in terms  of the
physical  variables $\vec{q}$,  $\vec{p}$ and  $t$, eliminating  the variables
$X$, $\Xi$, $\vec{\pi}$ and $s$ to obtain \cite{NH},
\begin{equa}[2] \label{eq:nh}
&\dot{q_i} \ \ &=& \ \ \ \frac{p_i}{m} \ ,\\
&\dot{p_i} \ \ &=& \ \ \ -\nabla\phi - \xi p_i \ ,\\
&\dot{\xi} \ \ &=& \ \ \
\frac{1}{\tnh^2}\left(\frac{1}{\nth\kB T}\sum_{i=1}^{\nth} \frac{p_i^2}{m} -
  1\right) \ ,\\
\end{equa}
where $\tnh$ is the relaxation time of the thermostat related to its effective
mass $Q$ as $\tnh^2 = Q/\nth\kB T$ and $\xi = \Xi/Q$, (in what follows we will
refer to the variable  $\xi$ as the thermostat).  Inspecting Eqs.~\eref{eq:nh},
known as the \NH equations of motion, the role of the thermostat $\xi$ becomes
evident:  it  acts  as  a  friction term  that  appropriately  changes  sign,
depending  on   whether  the  instantaneous  kinetic  energy   of  the  $\nth$
thermostated \dof  is larger or smaller  than the desired  mean kinetic energy
$\nth\kB T/2$.

To simulate a  nonequilibrium state, we couple \NH  thermostats to the angular
velocity of each of the four disks at the boundaries of the channel. Since the
disks are  pinned to  their positions, the  thermostats can interact  with the
rest  of the  \dof only  via  the collisions  with the  point-like moving  gas
particles.  We denote  by $\xiL^+$ and $\xiL^-$ the  thermostats for the upper
and lower disks at the cold  side and as $\xiR^+$ and $\xiR^-$ the thermostats
for the upper and lower disks at  the hot side, thus $\nth=1$. Note that it is
possible to couple the two disks at each boundary to one single thermostat, in
which  case, $\nth=2$.   We have  chosen the  former coupling  as we  get more
efficient numerics than  in the latter case. Also, we assume  that there is no
external potential $\phi$ acting on the particles.

In \cite{klages}, it has been shown  that the \NH dynamics are not ergodic for
$\phi=0$,  thus the  phase space  probability distribution  is  not canonical.
However, as we  will see, the interaction of the  thermostated disks with the
gas of particles leads to ergodic motions in the accessible phase space.

Labeling the cells of the channel from $1$ to $L$, the equations of motion for
our model can be written as follows
\begin{equa}[4] \label{eq:NH}
&\dot{q}_i  \ \ &=& \ \ \  \frac{p_i}{m} \quad &;& \quad i = 1,\ldots,n\\
&\dot{p}_i  \ \ &=& \ \ \  \Upsilon_{\eta;i}(t) \quad &;& \quad i = 1,\ldots,n\\
&\dot{\omega}_j^\pm  \ \ &=& \ \ \  \Upsilon_{\eta;j}^\pm(t)\quad &;& \quad j =
2,\ldots,L-1\\
\\
&\dot{\omega}_\L^\pm \ \ &=& \ \ \ \Upsilon_{\eta;\L}^\pm(t) - \xi_\L^\pm\omega_\L^\pm \ ,\\
&\dot{\omega}_\R^\pm \ \ &=& \ \ \ \Upsilon_{\eta;\R}^\pm(t) - \xi_\R^\pm\omega_\R^\pm \ ,\\
&\dot{\xi}_\L^\pm \ \ &=& \ \ \ \frac{1}{\tnh^2}\left(\frac{\Theta{\omega_\L^\pm}^2}{\kB T_\L} -  1\right) \ ,\\
&\dot{\xi}_\R^\pm \ \ &=& \ \ \ \frac{1}{\tnh^2}\left(\frac{\Theta{\omega_\R^\pm}^2}{\kB T_\R} -  1\right) \ ,\\
\end{equa}
where   $n$   is   the   number   of   particles.   The   abstract   operators
$\Upsilon_\eta(t)$,  represent the instantaneous  forces exerted  on particles
and disks during a collision. They are subjected to the condition
\begin{equ} \label{eq:constraint}
\sum_{i=1}^n \frac{p_i}{m}\Upsilon_{\eta;i}(t) +
\sum_{\substack{i=2\\\{+,-\}}}^{L-1}\Theta \omega_i\Upsilon_{\eta;i}(t)
+
\sum_{\{+,-\}}\Theta\left(\omega_\L\Upsilon_{\eta;\L}(t)+\omega_\R\Upsilon_{\eta;\R}(t)\right)
= 0 \ ,
\end{equ}
which accounts for the conservation of energy during the collisions.

The dynamics \eref{eq:NH}  is dissipative as the phase  space volume contracts
at a rate given by
\begin{equ} \label{eq:pscr}
\Lambda \equiv -\mathrm{div}_\Gamma \dot{\Gamma}=\xi_\L^-  + \xi_\L^+ + \xi_\R^-  + \xi_\R^+ \ ,
\end{equ}
where $\Gamma  \equiv (\vec{q},\vec{p},\vec{w},\vec{\xi})$. The  mean value of
the phase space  contraction rate $\Lambda$ is zero  only at equilibrium, {\em
  i.e.}, when  $T_H =  T_C$.  As we  will see in  Section~\ref{sec:Omega}, the
cold thermostats  contract the phase  space volume at  a rate faster  than the
expansion produced  by the  hot thermostats so  that on average,  $\Lambda$ is
positive.

\subsection{Ergodicity and Equilibrium State}
\label{sec:equil}

\begin{figure}[!t]
\begin{center}
  \includegraphics[scale=1.0]{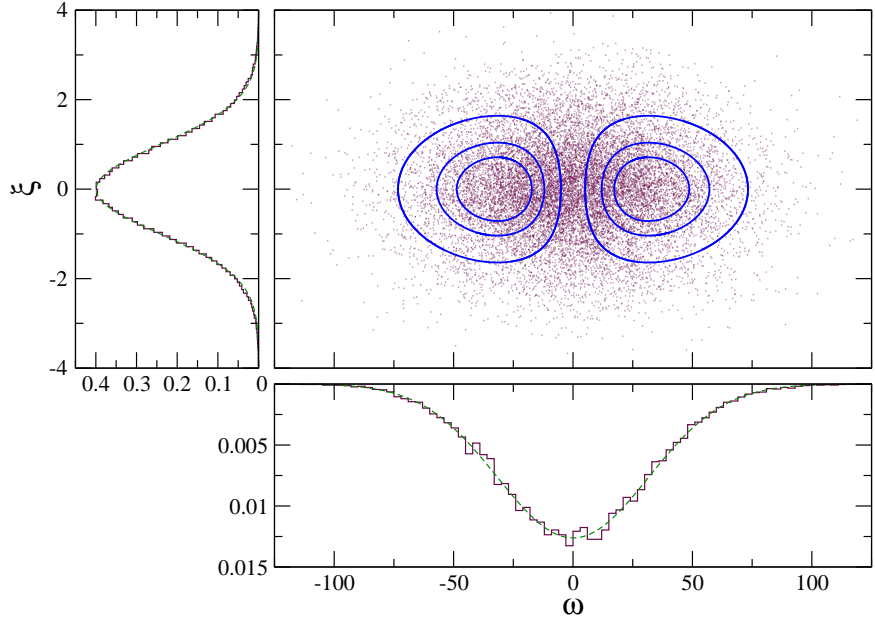}
\caption{Main panel: Evolution of the angular velocity of the disk $\omega(t)$
   and  its thermostat  $\xi(t)$  for the  one  single cell  at a  temperature
   $T=1000$, in the  presence (dots) and absence (solid curves)  of the gas of
   particles. In the presence of the particles the trajectory is obtained from
   one single initial condition. The solid curves correspond to three pairs of
   symmetric initial conditions.  Secondary panels: histograms of the marginal
   distributions  $P(\xi)$ (left)  and $P(\omega)$  (lower) of  the trajectory
   with   interactions.   The  dashed   curves  represent   the  corresponding
   equilibrium distributions.
\label{fig:ergo}}
\end{center}
\end{figure}

Because our  system is  not subjected to  any external field,  the thermostats
dissipate energy only when they collide with the particles.  These collisions,
in  turn, make  apparently  ergodic  the evolution  of  the thermostated  {\em
d.o.f.}, since the particles dynamics is  randomized by the motion in the bulk
of the system.  To illustrate this, we have considered the dynamics inside one
single  hexagonal cell  (see Fig~\ref{fig:model}),  containing  a thermostated
disk  and   $n=10$  particles,   with  reflecting  boundary   conditions.   In
Fig.~\ref{fig:ergo},  the evolution of  the thermostat  $\xi$ and  the angular
velocity of the disk $\omega$ is shown for a temperature $T=1000$. In the main
panel,  the  solid  curves  correspond   to  the  evolution  of  the  isolated
thermostated  disk,  {\em i.e.},  without  the  interaction  with the  gas  of
particles, for three pairs of symmetric initial conditions.  In this case, the
evolution  of the  vector $(\omega,\xi)$  is oscillatory:  $\omega$ oscillates
around  its equilibrium value  $\sqrt{T}$, preserving  its initial  sign.  The
thermostat  $\xi$  oscillates symmetrically  around  zero.  Differently,  when
particles are allowed to interact with the disk, the trajectory of any initial
condition explores  all the available  phase space, as  shown by the  cloud of
points in the main  panel of Fig.~\ref{fig:ergo}.  Furthermore, the histograms
in the left  and lower panels correspond to the  marginal distributions of the
points of  the trajectory, which  coincide with the  corresponding equilibrium
distributions of $\xi$ and $\omega$ (dashed curves).

Summarizing, while in the absence of  interactions the dynamics of the disk is
trivial (periodic),  the interactions with the particles  lead to trajectories
which appear to explore all  the available $(\omega,\xi)$ space. This apparent
ergodicity,  combined  with  the  observation  that the  distribution  of  the
particles  velocities  approximates  a  Maxwellian  at  the  temperature  $T$,
indicates  that  locally  (in the  cell)  a  form  of thermal  equilibrium  is
established.

\begin{figure}[!t]
\begin{center}
  \includegraphics[scale=1.0]{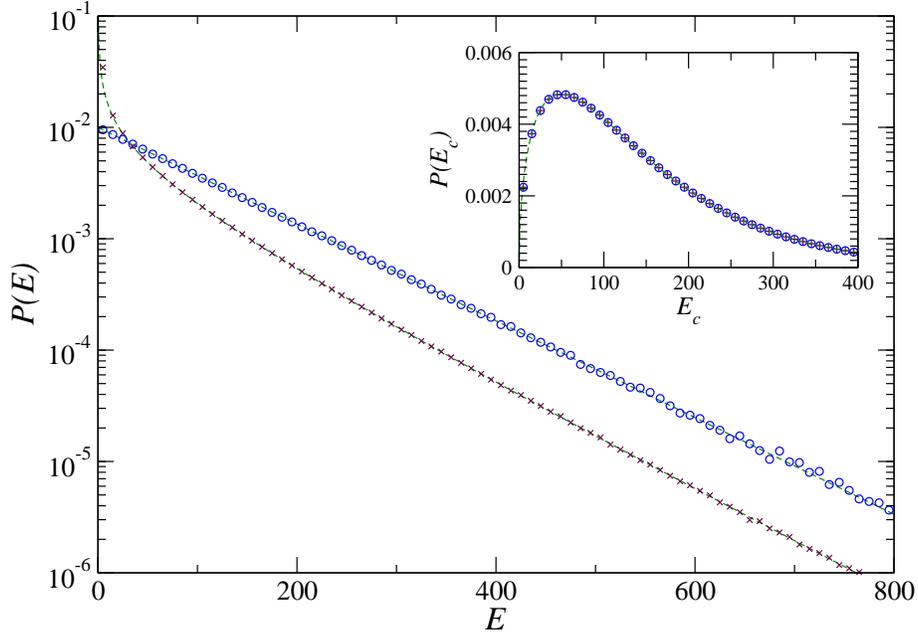}
\caption{  Equilibrium energy  probability  distribution $P(E)$  for the  disk
  (crosses) and particles (circles).  The distributions were obtained for $10$
  particles in a  hexagonal cell containing one single  disk thermostated at a
  temperature  $T=100$.   The  dashed  lines  correspond  to  the  theoretical
  equilibrium  distributions. Inset:  Conditional probability  distribution of
  the energy of the particles  $P(E_c)$, before (circles) and after (plus) the
  collisions.   The dashed  line corresponds  to the  theoretical distribution
  function.
\label{fig:dist-eq}}
\end{center}
\end{figure}

In the single  cell we also have computed  the energy probability distribution
function $P(E)$ of the  disk and of the energy of the  gas of particles, for a
given temperature  $T=100$. The  results are shown  in Fig.~\ref{fig:dist-eq}.
The  agreement with  the  theoretical equilibrium  Boltzmann distributions  is
excellent.  Independently of  the initial state, the gas  of particles and the
thermostated disk reach quite rapidly ($\mathcal{O}(10^{4})$ collisions times)
the equilibrium state  at a temperature equal to that  of the thermostat.  For
the  model depicted  in Fig.~\ref{fig:model}  a uniform  equilibrium  state is
established if $T_\L=T_\R=T$.

A  similarly  fast  relaxation  to  equilibrium, due  to  the  interaction  of
Eq.~\eref{eq:col-rules},   was   also    observed   using   stochastic   baths
\cite{mejia,larralde}.  The  local equilibration between  the thermostated and
not thermostated \dof is due to the fact that, at the collisions, the incoming
and outgoing temperatures dynamically relax to the same value. In the inset of
Fig.~\ref{fig:dist-eq}   the   incoming   and  outgoing   energy   probability
distribution function conditioned  to the occurrence of a  collision is shown.
Both PDF  coincide. Moreover, they  coincide with the  theoretical equilibrium
conditional probability (shown as a solid line).  This is a remarkable feature
of  the  present  model  which   is  not  always  obtained  in  other  systems
\cite{hoover04}.

Note  that  the  same  qualitative  results are  found  independently  of  the
particle's density $n$.  However, the  time required to reach equilibrium with
the gas  of particles is determined  by the rate  $\gamma\propto n\sqrt{T}$ at
which the  gas of particles collides  with the thermostated  disk, compared to
the relaxation time $\tnh$ of the thermostat.

Also note that the thermostated \emph{d.o.f.} $\omega$ changes discontinuously
at  collision, while  the  thermostat variable  $\xi$  does not.  We have  not
observed any  noticeable consequence of  the uncoupled evolution of  $\xi$ and
$\omega$ at the collisions.  We expect that this will affect the efficiency of
the thermostat, at most.  A similar observation was made in \cite{klages}, for
a thermostat with same kind of discontinuities.

\subsection{Nonequilibrium Steady State}
\label{sec:ness}

We now  turn our attention to nonequilibrium  states. In \cite{mejia,larralde}
the transport  properties of the system of  Sec.~\ref{sec:model} were studied.
The  system   was  coupled  to  stochastic  thermochemical   baths  to  create
temperature and  chemical potential gradients.  In that  configuration, it was
found  that the hypothesis  of local  thermal equilibrium  holds and  that the
deviations from it  are consistent with the transport of  heat and matter that
appears as a response to the  external gradients.  LTE results from the mixing
properties of  the effective interaction  among particles.  In the  absence of
interaction,  {\it  e.g.}, when  $\eta=0$,  LTE  is  not satisfied  (see  {\em
e.g.},\cite{dhar}).

In   our  present  situation,   the  stochastic   baths  are   substituted  by
deterministic \NH  thermostats. Moreover, the Lorentz channel  is closed, {\it
i.e.},  there  is  no  exchange  of  particles  between  the  system  and  the
reservoirs, implying that the particle current is zero.

\begin{figure}[!t]
\begin{center}
  \includegraphics[scale=1.0]{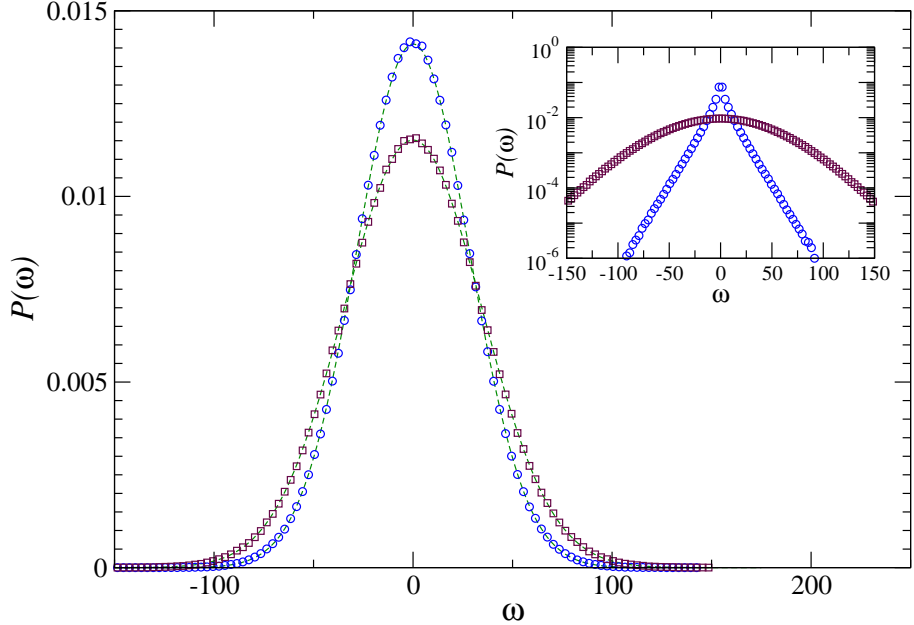}
\caption{
  Probability distribution function of the angular velocity $P(\omega)$ of the
  ``cold'' (circles)  and hot (squares) thermostated disks,  from a simulation
  with $T_C=900$  and $T_H=1100$, $n=20$ particles  per cell and  $L=7$ cells. 
  The  dashed lines  correspond  to the  theoretical equilibrium  distribution
  functions.   In the  inset, the  same distributions  are shown,  for  a much
  larger temperature gradient of $T_C=100$ (circles) and $T_H=1900$.
\label{fig:dist-noneq}}
\end{center}
\end{figure}

Since LTE  is a  consequence of the  mixing due  to the interactions,  we have
observed  that it holds,  indeed, in  the bulk.\footnote{By  bulk we  mean all
cells   which  do   not   contain   a  thermostated   disk   (open  disks   in
Fig.~\ref{fig:model}.}    However,  away   from  equilibrium,   this   is  not
necessarily the case  in the thermostated regions. In  order to understand how
effectively  the  interactions  establish  LTE  in the  neighbourhood  of  the
thermostated  disks, we  have  performed extensive  numerical simulations,  at
different  temperature gradients.   In Fig.~\ref{fig:dist-noneq}  we  show the
probability distribution function of  the angular momentum of the thermostated
disks $P(\omega)$, for a temperature gradient of $T_\C = 900$ and $T_\H=1100$.
The circles and the squares correspond to the distribution of the disks at the
cold  and  hot  sides,  respectively.   The dashed  lines  correspond  to  the
theoretical  equilibrium  distributions at  the  nominal  temperatures of  the
thermostats  $T_\C$  and $T_\H$.   The  agreement  of  $P(\omega)$ with  their
respective theoretical distributions is excellent.

However, for very  large temperature gradients, LTE is  not established at the
cold side,  while the LTE at the  hot side does not  necessarily correspond to
the chosen $T_H$.  For  instance, the inset of Fig.~\ref{fig:dist-noneq} shows
the  cold  and  hot  $P(\omega)$  for  $T_\C  =  100$  and  $T_\H=1900$.   The
exponential tails  of $P(\omega)$ at the cold  side is a signature  of the bad
equilibration.  While  at the hot  side $P(\omega)$ remains Gaussian,  but its
variance  corresponds  to   a  temperature  which  is  not   $T_H$.   The  bad
equilibration  around  the  cold  side  is  due to  the  inefficiency  of  the
thermostat to thermalize the much more energetic particles coming from the hot
region. Thus,  the bulk  of the system  practically equilibrates with  the hot
thermostat  alone.   This problem  can  be  partially  solved by  varying  the
relaxation time of the cold thermostat $\tnh$.  However, the state at the cold
side is very sensitive to variations  of $\tnh$, which must be determined case
by case.  In the following we  restrict our computations  to gradients $\Delta
T/T \le 1$, for which LTE, at the correct temperatures is established.

\begin{figure}[!t]
\begin{center}
  \includegraphics[scale=1.0]{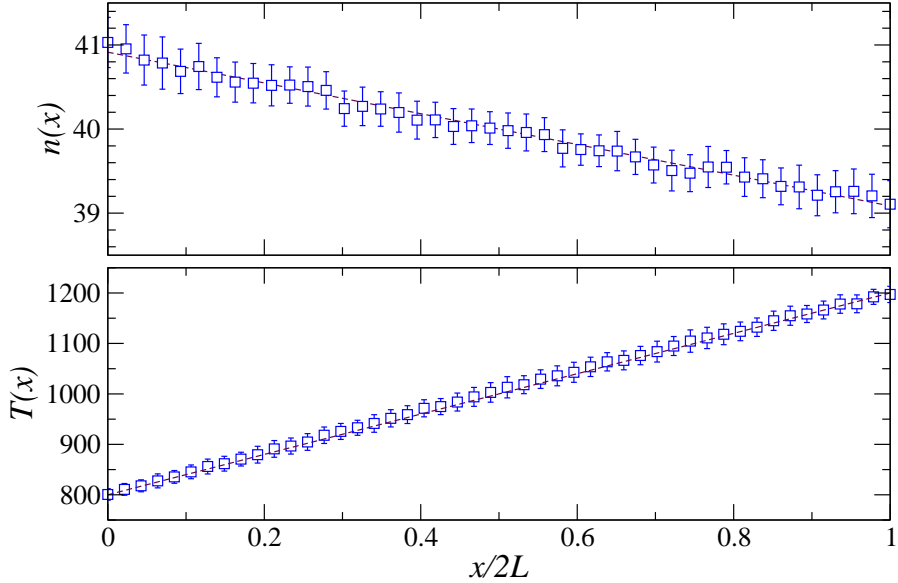}
\caption{ Stationary  profiles of  the density $n(x)$  and the  energy density
  $T(x)$ of  the particles  from a simulation  with $T_C=800$  and $T_H=1200$,
  $n=10$ particles per cell and $L=48$ cells. The dashed lines correspond to a
  linear profile between $T_C$ and $T_H$,  (lower panel), and to a best linear
  fit  of $n(x)$,  (upper  panel). Note,  however, that  for our  system
  neither $T(x)$ nor $n(x)$ need to be linear.
\label{fig:profiles}}
\end{center}
\end{figure}
In all  cases we have found that  the steady state is  characterized by linear
particle    $n(x)$     and    energy    $T(x)$     density    profiles.     In
Fig.~\ref{fig:profiles}, we show  an example of $n(x)$ and  $T(x)$ for a chain
of  $48$ cells and  $10$ particles  per cell,  with $T_C=800$  and $T_H=1200$.
Note that the  temperature gradient generates a particle  density gradient, in
the bulk.  However,  since the billiard model that we  consider is closed, the
mean particle  current is zero.   Note also that  $T(x)$ does not jump  at the
boundaries, which means that a  good equilibration between thermostats and gas
particles is achieved. By measuring the heat current as a function of the size
of the system, for a fixed temperature difference, we verified the validity of
Fourier's law to high accuracy.

The same  results had been  previously obtained in  \cite{mejia,larralde}, for
the system coupled  to stochastic thermochemical baths. It  is indeed expected
that the transport properties  in the bulk of the system do  not depend on the
particular model of heat bath.

\section{Fluctuation Relation for Heat Conduction}
\label{sec:ft}

In this section we study the fluctuations of the phase space contraction rate
and of the entropy production, or energy dissipation in the transient response
to external drivings and in the stationary state of the model system
introduced in Sec.~\ref{sec:model}.

\subsection{Transient Fluctuation Relation}
\label{sec:tft}

Consider a system that is initially at equilibrium. At time $t=0$, an external
force, either mechanical  or thermodynamical, is applied. For  times $t>0$ the
system evolves from its  equilibrium state towards, possibly, a nonequilibrium
steady  state.   In the  process,  the system  dissipates  the  work that  the
external  forcing  performs  on  it.   Furthermore, consider  an  ensemble  of
trajectories in  the phase space  $\Gamma$, that originate at  the equilibrium
state (at $t=0$) and  evolve for a finite time $\tau$.  For  $\delta > 0$, let
$A^+_\delta      =       (A-\delta,A+\delta)$      and      $A^-_\delta      =
(-A-\delta,A+\delta)$. The Transient  Fluctuation Relation \cite{ES02}, states
that
\begin{equ} \label{eq:tft}
  \ln\frac{\mathrm{Prob}\left(E(\overline{\Omega}_{0,\tau}\in
  A^+_\delta)\right)}{\mathrm{Prob}\left(E(\overline{\Omega}_{0,\tau}\in
  A^-_\delta)\right)}\simeq A\tau \ ,
\end{equ}
where  $E(\overline{\Omega}_{0,\tau}  \in (a,b))$  is  the  set  of points  in
$\Gamma$,    such   that    $\overline{\Omega}_{0,\tau}    \in   (a,b)$
\footnote{In what follows,  we will omit the specification of  the set $E$ for
  the             sake             of            simplicity.}              and
  $\overline{\Omega}_{0,\tau}=(1/\tau)\int_0^\tau  \Omega(\Gamma(t))dt$ is the
  time averaged {\em  dissipation function}, which is defined  in terms of the
  (initial)  phase  space probability  density  $f$  and  of the  phase  space
  contraction rate $\Lambda$ as \cite{ES02}:
\begin{equ} \label{eq:omega}
\overline{\Omega}_{0,\tau}(\Gamma) \equiv
\frac{1}{\tau}\left(\ln\frac{f(\Gamma)}{f(\mathcal{S}^\tau\Gamma)} +
\int_0^\tau \Lambda(\Gamma(t)) dt\right)
    \ .
\end{equ}

\begin{remark}
  The density $f$ can be arbitrarily chosen.  However, if it represents the
  equilibrium ensemble associated with the chosen dynamics, $\Omega$
  corresponds to the physical dissipation (the entropy production, close to
  equilibrium) \cite{SRE}.
\end{remark}

In this section we study the fluctuations of the time averaged dissipation
function as the system moves away from a given equilibrium state.  In
Ref.\cite{ES-heat}, a transient fluctuation formula for heat conduction was
obtained, considering a generic system coupled to two (deterministic)
thermostats, which is the case of the present work.  In Ref.\cite{SRE} the
transient fluctuation relation has then been obtained in great generality for
$\zW$, under the sole assumption of time reversibility. Hence we recall the
definitions of \cite{SRE}, which will be used for the steady state fluctuation
relations as well.

Consider a dynamical system with phase-space $\mathcal{M}$, described by an
evolution operator $\mathcal{S}^t~:~\mathcal{M}\rightarrow\mathcal{M}$.  Let
$\mathcal{S}^t$ be reversible, i.e.\ such that for an involution
$i~:~\mathcal{M}\rightarrow\mathcal{M}$ representing the time inversion
operator, $i\mathcal{S}^t\Gamma = \mathcal{S}^{-t}i\Gamma$ holds for all
$\Gamma\in\mathcal{M}$ and all $t\in\mathbb{R}$.  Introduce a probability
measure $\mu$ with density $f$ on $\mathcal{M}$, i.e.\ $d\mu(\Gamma) =
f(\Gamma)d\Gamma$, which is even for $i$, i.e.\ $f(i \zG)=f(\zG)$.  Let
$\phi~:~\mathcal{M}\rightarrow\mathbb{R}$ be any odd observable, {\em i.e.},
$\phi(i\Gamma) = -\phi(\Gamma)$, of a many-body dynamical system that
satisfies the assumptions above and denote by
\begin{equ}
\overline{\phi}_{0,\tau} = \frac{1}{\tau}\int_0^\tau \phi(\mathcal{S}^t\Gamma)
dt \ ,
\end{equ}
the time average of $\phi$ during a time $\tau$ with initial condition $\zG$.

In \cite{SRE} it was proved that $\phi$ satisfies the following symmetry
relation:
\begin{equ} \label{eq:phi-FR}
\frac{\mu\left(\overline{\phi}_{0,\tau} \in
    A^+_\delta\right)}{\mu\left(\overline{\phi}_{0,\tau} \in
    A^-_\delta\right)} =
    \langle e^{-\Omega_{0,\tau}}\rangle^{-1}_{\overline{\phi}_{0,\tau}\in
    A^+_\delta} \ ,
\end{equ}
where $\Omega_{0,\tau}(\Gamma) = \tau\overline{\Omega}_{0,\tau}(\Gamma)$ is the
time integral of $\zW$ over the interval $[0,\tau]$.  In the {\em r.h.s.}  of
\eref{eq:phi-FR} the average $\langle\cdot\rangle_{\overline{\phi}_{0,\tau}\in
A^+_\delta}$ is the ensemble average over the set of trajectories that satisfy
the constraint that $\overline{\phi}_{0,\tau} \in A^+_\delta$.  The
Conditional reversibility theorem derived in Ref.\cite{GG-patt} is related to
the steady state version of (\ref{eq:phi-FR}), given in \cite{SRE}.

When the system is subjected to an external mechanical force $\vec{F}$, the
density function leading to the physical dissipation is the equilibrium
($\vec{F}=0$) probability density.  Other choices are possible and, for
instance, the uniform density in a compact phase space,
$f(\Gamma)=1/\mathcal{|M|}$, yields $\Omega = \Lambda$, the phase space
contraction rate.

\begin{remark}
  The relation \eref{eq:phi-FR} is exact. It is purely a consequence of the
  time reversibility of the dynamics. Remarkably, this relation is valid for
  any observable that is odd with respect to time reversal.
\end{remark}
In particular, we can take $\overline{\Omega}_{0,\tau}$ as the observable.
Doing so, Eq.~\eref{eq:phi-FR} becomes
\begin{equa}[3] \label{eq:omega-FR}
& \frac{\mu\left(\overline{\Omega}_{0,\tau} \in
    A^+_\delta\right)}{\mu\left(\overline{\Omega}_{0,\tau} \in
    A^-_\delta\right)} &~ =~ & \langle e^{-\tau\overline{\Omega}_{0,\tau}}\rangle^{-1}_{\overline{\Omega}_{0,\tau}\in
    A^+_\delta}\\
& &~ =~ & e^{\left[A + \varepsilon(\delta,A,\tau)\right]\tau} \ , \\
\end{equa}
where the error term $\varepsilon\le\delta$, in general will depend on
$\delta$, $A$ and $\tau$.  In \cite{SRE} the transient fluctuation relation
\eref{eq:omega-FR} was called \OFR.

\subsection{Dissipation Function for Heat Flow}
\label{sec:Omega}

In this section we investigate the transient response of the system of
Sec.~\ref{sec:model} to a temperature gradient.  The dynamics of the model is
given by the equations of motion \eref{eq:NH} and the energy density of the
system $H$, can be written as
\begin{equa}[2] \label{eq:H0}
&H(\vec{p},\vec{\omega},\vec{\xi}) & \ = \ & \frac{1}{2m}\sum_{i=1}^n p_i^2 +
\frac{\Theta}{2}\sum_{j=1}^{L}\left[(\omega_j^+)^2 + (\omega_j^-)^2\right] + \\
& & \ + \ & \frac{\tnh^2}{2}\sum_{\{+,-\}}\left(T_\L\xi_\L^2 +
  T_\R\xi_\R^2\right) \ ,\\ 
\end{equa}
where $\omega_1^\pm = \omega_\C^\pm$ and $\omega_L^\pm = \omega_\H^\pm$. Here,
$\xiL = \xiL^- + \xiL^+$ and $\xiR = \xiR^- + \xiR^+$.  We consider the
following transient process:

At times $t<0$ the system and the thermostats are in thermal equilibrium at
some temperature $T_0$.  For simplicity we choose the equilibrium temperature
as $T_0=(T_\C+T_\H)/2$ (in the linear regime $T_0$ is the mean temperature of
the steady state of the system).  Therefore, the initial density distribution
corresponds to the canonical equilibrium distribution
\begin{equ} \label{eq:f}
f(\Gamma)  = \frac{e^{-\beta_0H(\Gamma)}}{\int d\Gamma e^{-\beta_0H(\Gamma)}} \ ,
\end{equ}
with $\beta_0 = 1/\kB T_0$, and $H(\Gamma)$ the energy density \eref{eq:H0},
with $T_\C=T_\H=T_0$.  At time $t=0$ the temperatures of the thermostats are
set to $T_\C$ for the {\em cold} thermostat and $T_\H$ for the {\em hot} one.
At times $t>0$ the system is not in equilibrium and a heat flux develops.

We measure the transient response of the system by considering time averages
during a finite time $\tau$ of the dissipation function $\zW$.

Substituting \eref{eq:f} into \eref{eq:omega}, the time average of
$\overline{\zW}_{0,\tau}$ takes the simple form
\begin{equ} \label{eq:omega-2}
\overline{\Omega}_{0,\tau}(\Gamma) = \frac{1}{\tau}\int_0^\tau\left(
\beta_0\dot{H}(S^t\Gamma) + \Lambda(S^t\Gamma)\right)dt =
\beta_0\left(\frac{H(S^t\Gamma)-H(\Gamma)}{\tau}\right) + \overline{\zL}_{0,\tau}(\Gamma) \ .
\end{equ}
The quantities $\zW$ and $\zL$, thus differ by a total time derivative, as in
the cases considered in e.g.\ \cite{ESR,romans,BGGZ,gilbert}.  Taking the time
derivative of \eref{eq:H0} and using the constraint \eref{eq:constraint} we
obtain

\begin{equ} \label{eq:Hdot}
\dot{H}(t) = -\left(T_\C\xi_\C + T_\H\xi_\H\right) \ .
\end{equ}

Finally, using the expression for the phase space contraction rate
\eref{eq:pscr}, the time averaged dissipation function for the heat flow of
our model becomes
\begin{equ} \label{eq:diss}
\overline{\Omega}_{0,\tau}(\Gamma) =
\frac{T_\H-T_\C}{T_\H+T_\C}\left(\overline{\xi}_{\C;~0,\tau} -
  \overline{\xi}_{\H;~0,\tau}\right) \ .
\end{equ}

\begin{figure}[!t]
\begin{center}
  \includegraphics[scale=1.0]{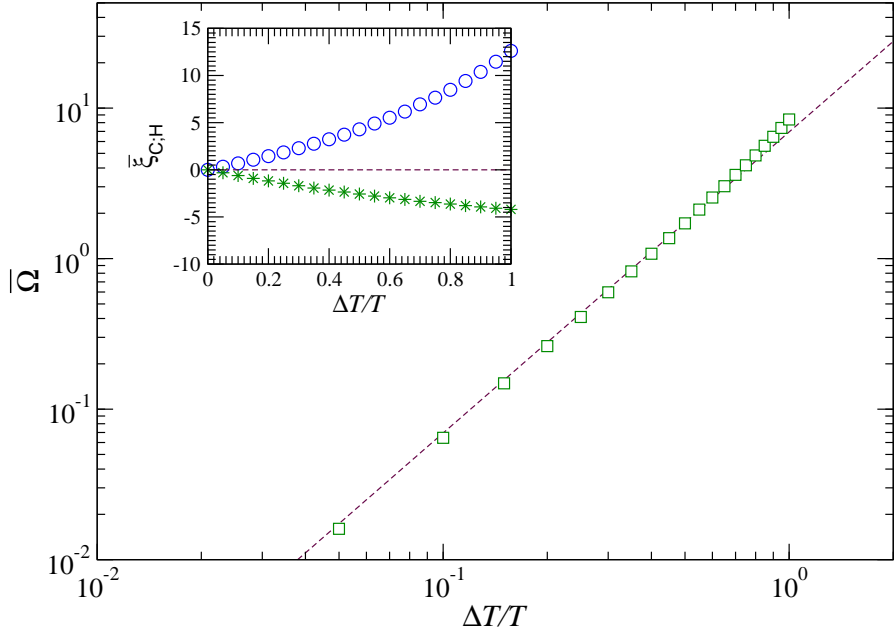}
\caption{ Long time average in the steady state of the dissipation function
  $\Omega = \frac{\Delta T}{T}(\xi_\C - \xi _\H)$ as a function of the
  relative temperature gradient $\Delta T/T$. The dashed line corresponds to
  $(\Delta T/T)^2$.  Inset: local phase space contraction rate,
  $\overline{\xi}_\C$ (circles), and $\overline{\xi}_\H$ (stars).
\label{fig:diss}}
\end{center}
\end{figure}

As a test of the physical character of the dissipation function
\eref{eq:diss}, we have computed its long time average in the steady state,
$\overline{\Omega}_{0,\tau}$, as a function of the relative temperature
gradient $\Delta T/T$.  This is shown in Fig.~\ref{fig:diss} for a chain of
$5$ cells and $20$ particles per cell.  For each simulation the temperatures
of the thermostats $T_\C$ and $T_\H$ were varied symmetrically around
$T=1000$.  At equilibrium $\overline{\Omega}_{0,\tau}=0$.  Out of equilibrium,
$\overline{\Omega}_{0,\tau}$ grows with the external gradient as $\left(\Delta
T/T\right)^2$.  In the inset of Fig.~\ref{fig:diss} we show the average
behaviour of the individual thermostats $\overline{\xi}_\C$ and
$\overline{\xi}_\H$.  At equilibrium, both thermostats average to zero. Away
from equilibrium, their mean values grow in absolute value, with the distance
from equilibrium.  While the dissipation at the cold thermostat contracts the
phase space volumes, the hot thermostat expands them.  However, $\xi_\C$
dissipates more than $\xi_\H$ creates, leading, on average, to a positive
global dissipation in the steady state.

To  test  the  Transient  Fluctuation  Relation  \eref{eq:omega-FR},  we  have
numerically computed the probability  distribution function of the dissipation
function \eref{eq:diss},  for different values  of the transient  time $\tau$,
and for  a channel of $L=10$ cells  with $n=4$ particles per  cell.  This test
may seem unnecessary,  because our dynamics is time  reversal invariant, which
is the only  condition required for (\ref{eq:omega-FR}) to  hold.  In reality,
there are two  subtle points, at least, which make  non-trivial the test: {\bf
a)}  our simulations  are not  exactly  reversible, because  of the  numerical
scheme approximating the  solution of the equations of  motion, and because of
round-off errors; {\bf b)} the initial continuous ensemble $f$, required by the
theory,  is replaced  by a  finite number  of initial  states, supposed  to be
picked up  at random with density $f$.  This, combined with the  fact that the
dissipation  takes place  discontinuously in  time,  because it  relies on  the
interactions between  thermostated disks  and point particles,  makes delicate
the choice of the numerical procedure, which could introduce spurious effects,
leading  to  fluctuations  that  do  not  satisfy  the  Transient  Fluctuation
Relation. Such spurious effects could spoil also the tests of the Steady State
Fluctuation  Relation  which,   differently  from  the  Transient  Fluctuation
Relation, requires more  than time reversibility to hold,  and certainly needs
to be tested case by case.

Therefore, the test of the Transient Fluctuation Relation is rather important,
in our  system, to  calibrate the numerical  protocols. Our  investigation has
shown that  the effect of the  numerical integration on  time reversibility is
negligible,  as far  as \eref{eq:omega-FR}  is  concerned, but  also that  the
construction of the initial ensemble needs special care.  What proved adequate
to  our purpose is  the construction  of $f$  from an  equilibrium simulation,
which  generates  a  collection  of  decorrelated  microstates,  by  saving  a
microstate every sufficiently long time.

We have proceed as follows: at $t=0$ the state of the system is set to a
microstate previously generated by the equilibrium dynamics at a temperature
$T_0=(T_\C+T_\H)/2$.  From this initial state, the temperatures of the
thermostats are set to $T_\C$ and $T_\H$ respectively.  Then, the system
evolved for a certain time $\tau$, and $\overline{\Omega}_{0,\tau}$ was
computed.  Considering a large ensemble of such processes, the empirical
probability distribution function $P(\overline{\Omega}_{0,\tau})$ was
constructed.

Furthermore, we define
\begin{equ} \label{eq:C-omega-def}
\mathcal{C}_\Omega(\tau;A) \equiv \frac{1}{\tau}\ln\frac{P\left(\overline{\Omega}_{0,\tau} \in
    A^+_\delta \right)}{P\left(\overline{\Omega}_{0,\tau} \in A^-_\delta\right)} \ , 
\end{equ}
where $A$ is the centre of the corresponding bin of the empirical
distribution. In terms of \eref{eq:C-omega-def}, the transient \OFR is written
as
\begin{equ} \label{eq:C-omega}
\mathcal{C}_\Omega(\tau;A) = A \ ; \ \textrm{for all} \ \ \tau \ .
\end{equ}

\begin{figure}[!t]
\begin{center}
  \includegraphics[scale=1.05]{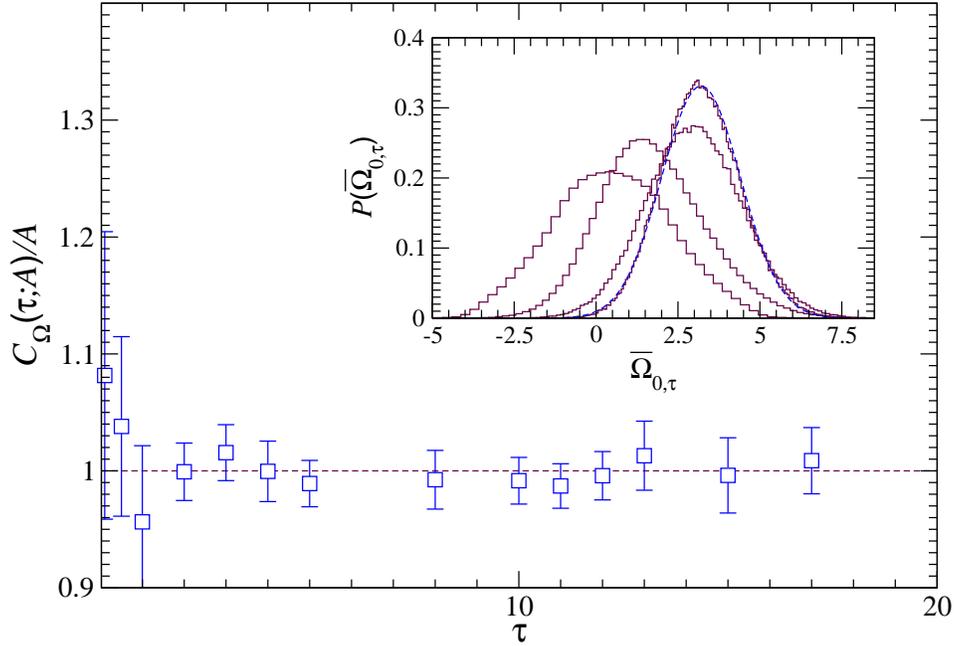}
\caption{ The slope of $C_\Omega(A,\tau)$ as a function of $\tau$, for a
  channel of $10$ cells, $T_\C = 700$ and $T_\H = 1300$.  Inset: Probability
  distribution function $P(\overline{\Omega}_{0,\tau})$ for (from left to
  right) $\tau=1$, $5$, $10$ and $15$. The dashed line corresponds to a
  Gaussian function.
\label{fig:tft-dists}}
\end{center}
\end{figure}

We have verified the Transient Fluctuation Relation for several values of
$\tau\in[0.01,20]$, for a temperature gradient given by $T_\C=700$ and
$T_\H=1300$. These results are shown in Fig.~\ref{fig:tft-dists} for the slope
of the relation \eref{eq:C-omega}.  In the inset of Fig.~\ref{fig:tft-dists}
we show $P(\overline{\Omega}_{0,\tau})$ for four values of $\tau$.  We have
found that for small $\tau$, the obtained distribution is highly
asymmetric. It becomes Gaussian only for large $\tau$.

The verification of \eref{eq:C-omega} indicates that our numerical procedure
is appropriate to test the $\zW$-FRs, hence that it may be used to test the
Steady State Fluctuation Relation.

\subsection{Steady State Fluctuation Relation}
\label{sec:SSFT}

The steady state fluctuation relation obtained by Evans, Cohen and Morriss in
\cite{ECM} was motivated by the fact that the phase space contraction rate
$\Lambda$ is proportional to the energy dissipation rate divided by the
temperature, for a subclass of Gaussian isoenergetic particle systems, which
includes the model of \cite{ECM}.  This way the fluctuation relation for
$\Lambda$, which we call \LFR as in \cite{SRE}, could be given a physical
meaning in \cite{ECM}. The \LFR has been later proved by Gallavotti and Cohen
to hold for all dissipative, time reversible, smooth and uniformly hyperbolic
dynamical systems, independently of their physical meaning \cite{GCa,GCb}.
For such systems, the \LFR states that:
\begin{equ} \label{eq:lfr}
\lim_{\tau\rightarrow\infty} \frac{1}{\tau \langle\Lambda\rangle_\infty}\log
  \frac{P(\hat{\Lambda}_\tau \in 
  A^+_\delta)}{P(\hat{\Lambda}_\tau\in A^-_\delta)} =
   A  \ , 
\end{equ}
where
$\hat{\Lambda}_\tau=\overline{\Lambda}_{0,\tau}/{\langle\Lambda\rangle}_\infty$
is the phase space contraction rate averaged during a time interval $\tau$ and
normalized to  its stationary state  value $\langle\Lambda\rangle_\infty$, and
$P(\hat{\Lambda})$ is the  corresponding steady state probability distribution
function.    The    \LFR    is     supposed    to    be    valid    only    if
$\langle\Lambda\rangle_\infty\ne  0$ and   $A$  belongs to  the  interval of
definition of $\langle\Lambda\rangle_\infty$.

In \cite{SRE}, it has been shown that the steady state fluctuation relations
may only rely on the time reversibility of the microscopic dynamics and on
the time behaviour of the quantity $\mathcal{B}_\zW$, defined in
Eq.~\eref{eq:omega-corr} below, independently of the fluctuating odd
observable under consideration.  This way, $\zW$ appears to be a sort of {\em
``thermodynamic potential''}, which determines the behaviour of all other odd
observables, but works even far from equilibrium.\footnote{In the mathematical
theory of smooth, uniformly hyperbolic dynamical systems, it is $\zL$ that
plays a role analogue to that of a ``thermodynamic potential''.}  This can be
understood starting from the symmetry relation \eref{eq:phi-FR}, concerning
the fluctuations of any observable that is odd with respect to time reversal.
In \eref{eq:phi-FR}, the symmetry of the fluctuations is expressed in terms of
averages of the dissipation function \eref{eq:omega}, with respect to the
initial measure $\mu$ of density $f$.  Taking the ensemble average with
respect to the measure evolved up to time $t$, $\mu_{t}$, of the averages of
$\zW$ along trajectory segments of duration $\tau$, which at time $t=0$
started from the initial state $\mu$, one obtains:
\begin{equ}[2] \label{eq:ssomega}
\frac{1}{\tau}\ln\frac{\mu_{t}\left(\overline{\Omega}_{0,\tau} \in
    A^+_\delta\right)}{\mu_{t}\left(\overline{\Omega}_{0,\tau} \in
    A^-_\delta\right)}  =  A + \epsilon(\delta,t,A,\tau) -
    \frac{1}{\tau}\ln\mathcal{B}_\Omega(t,\tau) \ ,
\end{equ}
where $\epsilon$ is bounded by $0 \le \epsilon<|\delta|$, and
$\mathcal{B}_\Omega(t,\tau)$ is a kind of correlation function given
by\footnote{Although the aspect of $\mathcal{B}_\Omega(t,\tau)$ resembles that
of a standard correlation function, it cannot be always interpreted like
that. For instance, in cases which verify the modified FR of axiom C systems,
which enjoy a strong form of correlations decay
\cite{BGG,BGaxiomC,RMaxiomC,GRS}, the term
$\tau^{-1}\ln\mathcal{B}_\Omega(t,\tau)$ does not decay, if $\zW$ is chosen
to equal $\zL$.  In other cases in which no FR is verified, such as those with
trivial steady states considered in \cite{LRB,CG99},
$\mathcal{B}_\Omega(t,\tau)$ grows exponentially with $t$, as simple
calculations show. Because the $t \to \infty$ limit must be taken before the
$\tau \to \infty$ limit, this illustrates that the transient FR cannot not
turn into a steady state FR, in these cases, although the transient FR holds
at all times.  A more detailed discussion of these facts can be found in
\cite{RMM,SRE}.}
\begin{equ}[2] \label{eq:omega-corr}
\mathcal{B}_\Omega(t,\tau) =  \langle e^{-\Omega_{0,t}}
e^{-\Omega_{t+\tau,2t+\tau}}\rangle_{\overline{\Omega}_{t,t+\tau}
\in A^+_\delta} \ .
\end{equ}
If $\mu_{t}$ tends to a steady state $\mu_\infty$ when $t \to \infty$, then
\eref{eq:ssomega} should change from a statement on the ensemble at time $t$,
to a statement on the statistics generated by a single typical trajectory of
the stationary state $\mu_\infty$.  This requires some assumption, because $t$
tends to infinity before $\tau$ does and, in principle, the growth of $t$
could make the conditional average in \eref{eq:omega-corr} diverge. However,
if a finite time scale $\tau_m$ characterizes the decay of $\tau^{-1} \ln \mathcal{B}_\zW$,\footnote{Note 
that $\mathcal{B}_\Omega$ is computed with respect to the initial measure $\mu$, even for $t>0$} 
then \eref{eq:ssomega} shows that $\tau_m$
is the time scale which must be exceeded for the $t \to \infty$ limit to be
practically reached. When this is the case, the transient relations yield the
steady state ones, in the long time limits \cite{SRE}.

In some  cases, the steady state  \LFR and \WFR coincide,  as mentioned above,
while in other cases they are  clearly different.  For instance, in some cases
in which $\zL$ represents the heat  flow, $\zW$ the dissipated energy, and the
particles interaction potentials are singular,  the tails of the \LFR and \WFR
may be quite different \cite{ESR,SRE,VZC,BGGZ}
\footnote{This behaviour was first obtained for a stochastic model, namely the
Brownian  particle  dragged  in  a  liquid  by  a  moving  harmonic  potential
\cite{VZC}.   Later, the  corresponding  extended \LFR  has  been derived  for
observables for  which the  probability distribution function  has exponential
tails \cite{BGGZ}.}.  Which systems  actually enjoy the decay of $\tau^{-1}\ln
\mathcal{B}_\Omega(t,\tau)$ required  for \eref{eq:lfr} to hold,  or for other
steady  state relations  to be  derived from  (\ref{eq:phi-FR}) is  not known.
Hence a test is necessary in the systems of physical interest
\footnote{It is known, however, that $\tau^{-1}\ln\mathcal{B}_\zW$ has to
decay in systems not too far from equilibrium, i.e.\ in the linear regime of
Irreversible Thermodynamics, for the transport coefficients to exist
\cite{RMM,SRE}.}.  Therefore, we perform the test on our model.

The mesoscopic time scale $\tau_m$, expresses a physical property of the
system, typically the mean free time and thus, does not depend on $t$ or
$\tau$.  If $t , \tau \gg \tau_m$, the boundary terms
$\overline{\Omega}_{t-\tau_m,t}$ and
$\overline{\Omega}_{t+\tau,t+\tau+\tau_m}$, are typically small compared to
$\overline{\Omega}_{t,t+\tau}$. For this not to be the case, some singularity
of ${\Omega}$ must occur within $(t-\tau_m,t)$ or $(t+\tau,t+\tau+\tau_m)$.
However, even in that case, similar events may equally likely occur in the
intervals $(0,t)$ and $(t+\tau,2t+\tau)$, hence
$\overline{\Omega}_{t-\tau_m,t}$ and
$\overline{\Omega}_{t+\tau,t+\tau+\tau_m}$ are expected to contribute only a
fraction of order $\mathcal{O}(\tau_m/\tau)$ to the arguments of the
exponentials in \eref{eq:omega-corr}.  Therefore, in the cases of
thermodynamic interest, one can write
\begin{equa}[2]\label{newargdeco}
& \mathcal{B}_\Omega(t,\tau) & \ \approx \ & 
\left\langle e^{-{\Omega}_{0,t-\tau_m}} \cdot e^{- {\Omega}_{t+\tau+\tau_m,2t+\tau}}
\right\rangle_{{\overline{\Omega}}_{t,t+\tau} \in A^+_\delta} \nonumber \\
& & \ \approx \ & \left\langle e^{-{\Omega}_{0,t-\tau_m}} \cdot e^{- {\Omega}_{t+\tau+\tau_m,2t+\tau}}
\right\rangle \nonumber \\
& & \ \approx \ & \left\langle e^{-{\Omega}_{0,t+\tau_m}} \right\rangle
\left\langle e^{- {\Omega}_{t+\tau+\tau_m,2t+\tau}} \right\rangle\\
& & \ \approx \ & \mathcal{O}(1) \ ,
\end{equa} 
with an accuracy which improves when $t$ and $\tau$ are growing multiples of
the time $\tau_m$.  The last line is just a statement of the conservation of
probability and of the fact that $\mathcal{B}_\zW$ typically decays within
microscopic times \cite{RMM}.  One then realizes that
$\mathcal{B}_\Omega(t,\tau)$ tends to approximately $1$ as $1/\tau$, with a
characteristic scale of order $\mathcal{O}(\tau_m)$, which leads to the
validity of the steady state \WFR.  Then, for sufficiently small $\delta$, and
large $t$ and $\tau$ the relation
\begin{equ} \label{eq:ssomega-2}
\frac{1}{\tau}\ln\frac{\mu_{t}\left(\overline{\Omega}_{t,t+\tau} \in
    A^+_\delta\right)}{\mu_{t}\left(\overline{\Omega}_{t,t+\tau} \in
    A^-_\delta\right)}  \approx  A \ .
\end{equ}
holds, and remains valid in the $t \to \infty$ limit.  If $P$ is the
corresponding steady state probability distribution function, one can rewrite
\eref{eq:ssomega-2} as
\begin{equ} \label{eq:ssomega-3}
\mathcal{C}_{\Omega_\infty}(\tau;A) \equiv
    \frac{1}{\tau}\ln\frac{P\left(\overline{\Omega}_{0,\tau} \in
    A^+_\delta\right)}{P\left(\overline{\Omega}_{0,\tau} \in
    A^-_\delta\right)}  \approx    A \ ; \  
    \textrm{for} \quad \tau > \tau_m \ ,
\end{equ}
which we call steady state $\zW$-FR.

We have measured the steady state fluctuations of $\Lambda$ and of $\Omega$,
for the model introduced in Section \ref{sec:model}.  Let us first describe
those of $\zL$, and then compare them with those of $\zW$, in the
$\tau\rightarrow\infty$ limit.

\begin{figure}[!t]
\begin{center}
  \includegraphics[scale=1.1]{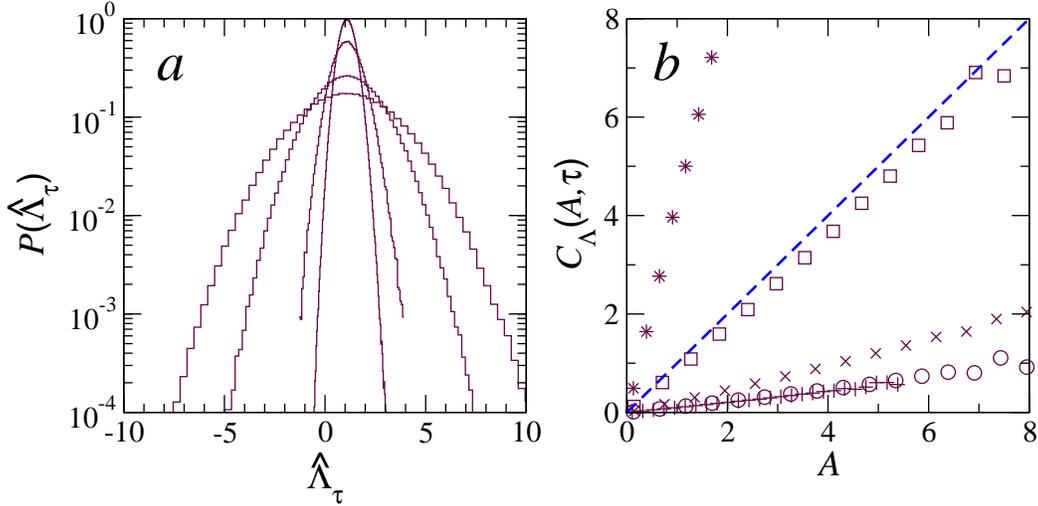}
\caption{Panel $a$:  Probability distribution $P(\Lambda_\tau)$  for $\tau=1$,
  $10$, $20$ and $40$. The largest the $\tau$ the more narrow the distribution
  gets.  The data was obtained for a channel of $L=10$ cells with a density of
  $n=4$ particles per  cell.  The temperatures of the  thermostats were set to
  $T_\C=700$ and $T_\H=1300$.  Panel $b$: $C_\Lambda(\tau;A)$ as a function of
  $A$  for different  values of  $\tau$: $0.1$  (stars), $0.5$  (squares), $1$
  (crosses), $5$ (circles) and $10$  (pluses).  The dashed line corresponds to
  the line with slope  $1$.  As observed, $C_\Lambda(\tau;A)$ is approximately
  linear  with $A$, for  all $\tau$,  with a  slope $\ne  1$, that  depends on
  $\tau$ (see Fig.~\ref{fig:convergence}  below).  Note that $P(\Lambda_\tau)$
  are not symmetric  about their maximum. Therefore, for  the values of $\tau$
  considered, $P(\Lambda_\tau)$ is not Gaussian, but their tails decay as fast
  as Gaussian.
\label{fig:ssft-vs-tau}}
\end{center}
\end{figure}

We consider a channel of length $L=10$ cells and $n=4$ particles per cell. As
before, the disks in the leftmost and rightmost cells are coupled to \NH
thermostats with $T_\C=700$ and $T_\H=1300$ respectively, and $\tnh=1$.
Starting from a random initial state we leave the system relax for a time
corresponding to $5\times10^6$ collisions of the particles with the disks.
This relaxation time ensures that the system has reached its stationary state
at which, the mean heat current is uniform.  Using Eq.~\eref{eq:pscr}, in the
stationary state we measure the values of the phase space contraction rate
$\Lambda$ each $10^{-3}$ time units.  We then compute the time average
$\overline{\Lambda}_{0,\tau}$, from disconnected intervals of the time series
of $\Lambda$.  Finally, we compute the empirical probability distribution
function of $\overline{\Lambda}_{0,\tau}$, normalized to its steady state
value $\langle\Lambda\rangle_\infty$.

We  have found  that, for  our  system, the  unboundness of  $\Lambda$ is  not
associated with  exponential tails in  its probability distribution.   This is
shown  in  Fig.~\ref{fig:ssft-vs-tau}-$a$,  where  $P(\hat{\Lambda}_\tau)$  is
plotted   for  different   values  of   $\tau$:  within   numerical  accuracy,
$P(\hat{\Lambda}_\tau)$  has Gaussian tails  for all  values of  $\tau$. Note,
however, that for the  values of $\tau$ considered, $P(\hat{\Lambda}_\tau)$ is
not a Gaussian distribution.

To discuss the behaviour of the \LFR \eref{eq:lfr} for finite averaging times
$\tau$, we define, as before,
\begin{equ}\label{eq:lslope}
\mathcal{C}_\Lambda(\tau;A) \equiv \frac{1}{\tau \langle\Lambda\rangle_\infty}
  \ln\frac{P(\hat{\Lambda}_\tau\in
  A^+_\delta)}{P(\hat{\Lambda}_\tau \in A^-_\delta)} \ ,
\end{equ}
so that the relation \eref{eq:lfr}), for finite but sufficiently large $\tau$,
would imply  that $\mathcal{C}_\Lambda(\tau;A)/A \approx 1$.  In  panel $b$ of
Fig.~\ref{fig:ssft-vs-tau}, the dependence of $\mathcal{C}_\Lambda(\tau;A)$ on
$A$ is  shown for different values  of $\tau$.  Independently of  the value of
$\tau$,  $\mathcal{C}_\Lambda(\tau;A)$ is  approximately linear  in  $A$.  The
only dependence  on $\tau$ is  the slope of  $\mathcal{C}_\Lambda(\tau;A)$ and
the  value of  $A$ at  which  the statistical  errors become  large.  This  is
different  from  recent  findings  for  other \NH  thermostated  systems.   In
\cite{gilbert} the \NH thermostated Lorentz  gas was studied, and it was found
that the  fluctuations of $\Lambda$ obeying  the \LFR are only  those that are
smaller      in     magnitude     than      the     steady      state     mean
$\langle\Lambda\rangle_\infty$. Larger  heat fluctuations follow  the extended
version of  the \LFR  of \cite{VZC,BGGZ}, which  saturates at some  value.  In
these works,  the deviation from the standard  \LFR was supposed to  be due to
the  singularities  of $\zL$,  which  may  produce  exponential tails  in  the
probability distribution function  of $\zL$, possibly as a  consequence of the
exponential tails of the distribution function of the values of $H$.

Our results confirm that  the identification of deterministic particle systems
which verify a modified version of  the FR is far from obvious \cite{ESR,SRE}.
Indeed,  the  most  common  situation   which  is  found  in  the  specialized
literature, as well  as here, is that even systems  with singular $\zL$ verify
the standard  FR, since the tails  of the distribution of  the fluctuations of
$\zL$  decay  as  fast  as  Gaussian  tails.   Then,  as  explained  e.g.\  in
Ref.\cite{BGGZ},  the standard FR  must hold.   In our  case, the  phase space
contraction rate  $\Lambda$ is unbounded because of  the logarithmic potential
present  in  the Nos\'e  Hamiltonian  \eref{eq:nose},  but  the tails  of  the
distributions   of    its   fluctuations   appear   to    be   Gaussian,   and
$\mathcal{C}_\Lambda(\tau;A)$  turns  out  to  be  linear  in  $A$,  even  for
fluctuations much larger than $\langle \zL \rangle_\infty$.

\begin{figure}[!t]
\begin{center}
  \includegraphics[scale=1.0]{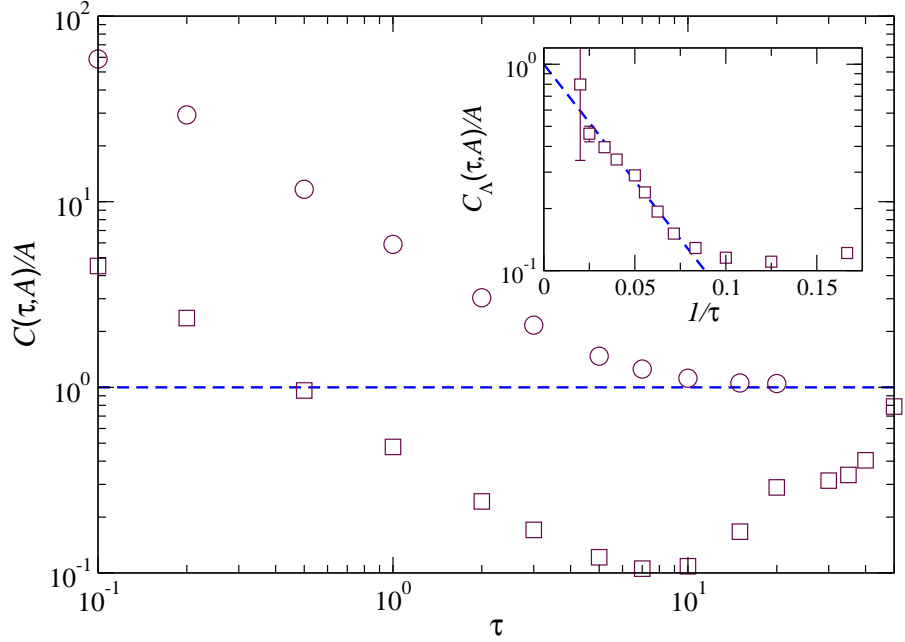}
\caption{ Comparison between the slopes of $\mathcal{C}_\Lambda(\tau;A)$
  (squares), and $\mathcal{C}_{\Omega_\infty}(\tau;A)$ (circles), as a
  function of the averaging time $\tau$.  The parameters are as indicated in
  Fig.~\ref{fig:ssft-vs-tau}.  In the inset, $\mathcal{C}_\Lambda(\tau;A)/A$
  is plotted in inverse units of $\tau$. The dashed line corresponds to the
  function $e^{-26/\tau}$.
\label{fig:convergence}}
\end{center}
\end{figure}

We now discuss the behaviour of $\mathcal{C}_\Lambda(\tau;A)$ with
$\tau$.  In Fig.~\ref{fig:convergence}, the slope of the
$\mathcal{C}_\Lambda(\tau;A)$ (squares), is shown as a function of $\tau$.
These values were obtained from a best fit to a linear function of
$\mathcal{C}_\Lambda(\tau;A)$.  For small times $\tau$, shorter than the
mesoscopic time scale $\tau_m$, the slope of $\mathcal{C}_\Lambda(\tau;A)$
decreases approximately as $\sim 1/\tau$.  The slope reaches a minimum value,
after which $\mathcal{C}_\Lambda(\tau;A)/A$ increases, seemingly converging to
its asymptotic value $1$.  This convergence becomes clearer in the inset
of Fig.~\ref{fig:convergence}, where the same data has been plotted as a
function of $1/\tau$.  These results are consistent with an exponential
convergence of $\mathcal{C}_\Lambda(\tau;A)/A$ to $1$ as
$\tau\rightarrow\infty$ (the dashed line in the inset of
Fig.~\ref{fig:convergence}, corresponds to an exponential scaling of
$\exp(-26/\tau)$).  Recalling Eq.(\ref{eq:Hdot}), and the fast convergence of
the $\zW$-FR, one may think that the monotonic convergence of the $\Lambda$-FR
goes as $1/\tau$, rather than exponentially.  Indeed, in the cases in which
$\Lambda$ is bounded, a conservative estimate of the convergence rate yields
$(\zL_{M}-\zL_m)/\tau$, where $\zL_{M}$ and $\zL_m$ are the maximum and
minimum values of $\zL$, see e.g.\ \cite{ESR}.  However, convergence may be
faster even when $\zL$ has no bounds, as in our case.  Therefore, the steady
state fluctuations of the phase space contraction rate for our model system
satisfy the $\Lambda$-FR \eref{eq:lfr}.

Let us now turn to the steady state \WFR \eref{eq:ssomega-3} and compare its
behaviour at large averaging times $\tau$ with that of the fluctuations of the
$\Lambda$-FR.  Fig.~\ref{fig:convergence} shows the slope of
$\mathcal{C}_{\Omega_\infty}(\tau;A)$ (circles), as a function of $\tau$.  For
times shorter than the mesoscopic time scale $\tau_m$, the steady state
fluctuation relations of $\Lambda$ and of $\Omega$ converge with an error
which decays as $1/\tau$.  At longer times $(\tau > \tau_m)$ the slope of
$\mathcal{C}_{\Omega_\infty}$ is fixed to $1$, indicating that the steady
state $\Omega$-FR \eref{eq:ssomega-2} holds.  Moreover,
Fig.\ref{fig:convergence} shows that the convergence of
$\mathcal{C}_{\Omega_\infty}$ is much faster than that of
$\mathcal{C}_\Lambda$, as expected because of the total derivative by which
$\zW$ and $\zL$ differ.  Indeed, we observed that the turning point of the
convergence of the \LFR occurs when $\mathcal{B}_\zW$ has decayed, and the
\WFR has converged, {\em i.e.}\ when only the term in $H$ is left to
decay. The difference in convergence rates is expected to grow when the system
approaches equilibrium, and the dissipative part of $\zL$ becomes dominated by
the total derivative $\dot{H}$.\footnote{Similar observations were made in
\cite{romans}, where the issue of convergence rates of FRs was thoroughly
investigated in models of shearing fluids. In the case of \cite{romans}, the
convergence of the $\zL$-FR appeared so much slower than that of the $\zW$-FR,
that it could not be directly observed, and was only indirectly inferred.}

\begin{figure}[!t]
\begin{center}
  \includegraphics[scale=1.0]{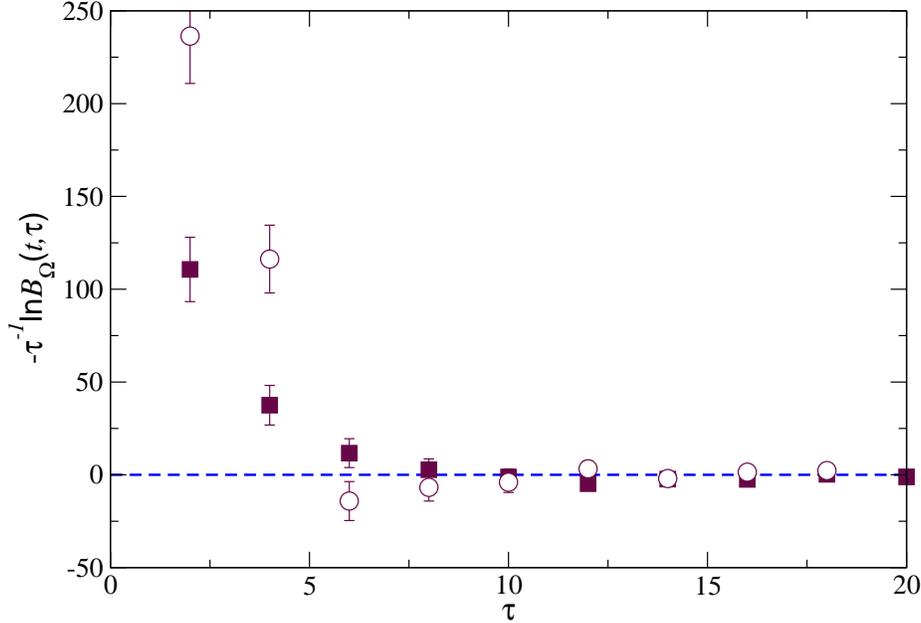}
\caption{Logarithm of $\mathcal{B}_\Omega(t,\tau)$ of Eq.~\ref{eq:omega-corr}
  as a function of the averaging time $\tau$, for $t=1$ (solid squares) and
  $t=10$ (open circles), for $A=2$ and $\delta=0.1$.  The parameters are as
  indicated in Fig.~\ref{fig:ssft-vs-tau}.
\label{fig:correlation}}
\end{center}
\end{figure}

These observations support the idea of \cite{SRE} that the scale set by the
decay of $\tau^{-1}\ln\mathcal{B}_\zW$ determines the behaviour of the
fluctuations of all odd observables, in our model, although the global
behaviour of different observables may be different. To test this we have
computed the behaviour of $\mathcal{B}_\zW(t,\tau)$. We considered a chain of
$L=10$ cells, density $n=4$ particles per cell and $T_\C=700$, $T_\H=1300$.
As in Section \ref{sec:Omega}, at $t=0$ the system is set to an equilibrium
microstate at temperature $T_0=(T_\C+T_\H)/2$. With the temperature of the
thermostats set to $T_\C$ and $T_\H$ respectively, we follow the evolution of
the system for a time $2t+\tau$ and compute the time averages of
$\Omega_{0,t}$, $\Omega_{t,t+\tau}$ and $\Omega_{t+\tau,2t+\tau}$.  Finally,
using Eq.~\eref{eq:omega-corr} and a large ensemble of such processes, we
obtain $\mathcal{B}_\zW(t,\tau)$.  Fig.~\ref{fig:correlation} shows the
dependence of $-\tau^{-1}\ln\mathcal{B}_\zW(t,\tau)$ on $\tau$ for $A=2$,
$\delta=0.1$, and $t=1$ (solid squares) and $t=10$ (open circles).  For both
values of $t$, $-\tau^{-1}\ln\mathcal{B}_\zW(t,\tau)$ decays, and reaches zero within
numerical accuracy, for times $\tau_m \approx 10$. This time scale coincides
approximately with the time scale at which, $\mathcal{C}_{\Omega_\infty}/A$
converges to its asymptotic value 1, and $\mathcal{C}_\Lambda/A$ starts its
monotonic convergence.

\section{Conclusions}
\label{sec:concl}

We have discussed the behaviour of the transient and steady state fluctuations
of the phase space contraction rate and of the dissipation function, in a \NH
thermostated system.  The dynamics in the bulk of the system is purely
Hamiltonian, hence preserves the phase space volumes. Two \NH thermostats, are
coupled to \dof that are fixed at the boundaries of the system.

We have studied the dynamics of the local thermostats and have characterized
the equilibrium and nonequilibrium states of the system. In the equilibrium
state, we have shown that the effective interaction among the particles is
sufficient to ergodize the, otherwise regular, dynamics of the thermostat.
This leads to a local thermalization of all the \dof of the system, for
$\Delta T / T < 1$ (which is our case).  Out of equilibrium, when the
temperatures of the thermostats are unequal, the system develops
nonequilibrium temperature and density profiles.  Consequently, a stationary
and uniform heat current appears, transporting heat from the hot to the cold
reservoir.  Energy is then dissipated at a constant rate, measured by the
dissipation function.

We have found that the transient fluctuations of $\zW$ do satisfy the
transient $\zW$-FR, indicating that our numerical solution of the equations of
motion and generation of the equilibrium ensemble from the equilibrium
dynamics are appropriate.

For the  system studied in this paper,  in which $\zL$ does  not coincide with
the energy  dissipation rate $\zW$,  the fluctuations of both  quantities obey
the  same   symmetry  given   by  the  standard   steady  state  FR,   in  the
$\tau\rightarrow\infty$  limit.   In  particular,  the  standard  $\zL$-FR  is
verified despite  the fact that $\zL$  is unbounded, because the  tails of the
distribution of  its fluctuations decay as  fast as Gaussian  tails.  The \LFR
needs  $\tau^{-1}\ln  \mathcal{B}_\zW$  to  have decayed,  before  starting  a
monotonic  convergence,  while the  \WFR  appears  to  have converged,  within
numerical accuracy,  after those times.   This different behaviour  depends on
the  total derivative $\dot{H}$,  which distinguishes  $\zL$ from  $\zW$.  The
characteristic time  scale of the  decay of $\tau^{-1}\ln  \mathcal{B}_\zW$ is
the mesoscopic time scale $\tau_m$.

\vskip 20pt

\noindent
{\bf  Acknowledgments:}   The  authors  are  indebted  to   D.J.\  Evans,  for
enlightening discussions and invaluable correspondence. We thank the anonymous
referees  for their  very detailed  and enlightening  remarks that  lead  to a
substantial improvement of  the paper.  We thank Fondazione  CRT for financial
support.

\end{document}